\documentclass[%
 reprint,
superscriptaddress,
 amsmath,amssymb,
 aps,
]{revtex4-2}

\usepackage{graphicx}
\usepackage{dcolumn}
\usepackage{bm}
\usepackage{siunitx}
\DeclareSIUnit\gl{\gram\per\litre}
\DeclareSIUnit\mev{\mega\eV}
\usepackage[version=3]{mhchem}
\usepackage{amsmath} 
\usepackage{caption}
\usepackage{subcaption}
\usepackage{float}
\usepackage[hidelinks]{hyperref}
\usepackage{xcolor}
\hypersetup{
    colorlinks,
    linkcolor={red!50!black},
    citecolor={blue!50!black},
    urlcolor={blue!80!black}
}

\begin{document}


\title{Event-by-Event Direction Reconstruction of \\Solar Neutrinos in a High Light-Yield Liquid Scintillator}

\author{ A.\,Allega}
\affiliation{\it Queen's University, Department of Physics, Engineering Physics \& Astronomy, Kingston, ON K7L 3N6, Canada}
\author{ M.\,R.\,Anderson}
\affiliation{\it Queen's University, Department of Physics, Engineering Physics \& Astronomy, Kingston, ON K7L 3N6, Canada}
\author{ S.\,Andringa}
\affiliation{\it Laborat\'{o}rio de Instrumenta\c{c}\~{a}o e  F\'{\i}sica Experimental de Part\'{\i}culas (LIP), Av. Prof. Gama Pinto, 2, 1649-003, Lisboa, Portugal}
\author{ J.\,Antunes}
\affiliation{\it Laborat\'{o}rio de Instrumenta\c{c}\~{a}o e  F\'{\i}sica Experimental de Part\'{\i}culas (LIP), Av. Prof. Gama Pinto, 2, 1649-003, Lisboa, Portugal}
\affiliation{\it Universidade de Lisboa, Instituto Superior T\'{e}cnico (IST), Departamento de F\'{\i}sica, Av. Rovisco Pais, 1049-001 Lisboa, Portugal}
\author{ M.\,Askins}
\affiliation{\it University of California, Berkeley, Department of Physics, CA 94720, Berkeley, USA}
\affiliation{\it Lawrence Berkeley National Laboratory, 1 Cyclotron Road, Berkeley, CA 94720-8153, USA}
\author{ D.\,J.\,Auty}
\affiliation{\it University of Alberta, Department of Physics, 4-181 CCIS,  Edmonton, AB T6G 2E1, Canada}

\author{ A.\,Bacon}
\affiliation{\it University of Pennsylvania, Department of Physics \& Astronomy, 209 South 33rd Street, Philadelphia, PA 19104-6396, USA}
\author{J.\,Baker}
\affiliation{\it SNOLAB, Creighton Mine \#9, 1039 Regional Road 24, Sudbury, ON P3Y 1N2, Canada}
\affiliation{\it Laurentian University, School of Natural Sciences, 935 Ramsey Lake Road, Sudbury, ON P3E 2C6, Canada}
\author{ N.\,Barros}
\affiliation{\it Laborat\'{o}rio de Instrumenta\c{c}\~{a}o e  F\'{\i}sica Experimental de Part\'{\i}culas (LIP), Av. Prof. Gama Pinto, 2, 1649-003, Lisboa, Portugal}
\affiliation{\it Universidade de Coimbra, Departamento de F\'{\i}sicai (FCUC), 3004-516, Coimbra, Portugal}
\author{ F.\,Bar\~{a}o}
\affiliation{\it Laborat\'{o}rio de Instrumenta\c{c}\~{a}o e  F\'{\i}sica Experimental de Part\'{\i}culas (LIP), Av. Prof. Gama Pinto, 2, 1649-003, Lisboa, Portugal}
\affiliation{\it Universidade de Lisboa, Instituto Superior T\'{e}cnico (IST), Departamento de F\'{\i}sica, Av. Rovisco Pais, 1049-001 Lisboa, Portugal}
% ...

\collaboration{The SNO+ Collaboration}


\begin{abstract}
\normalsize
\newpage
The direction of individual $^8$B solar neutrinos has been reconstructed using the SNO+ liquid scintillator detector. Prompt, directional Cherenkov light was separated from the slower, isotropic scintillation light using time information, and a maximum likelihood method was used to reconstruct the direction of individual scattered electrons. A clear directional signal was observed, correlated with the solar angle. The observation was aided by a period of low primary fluor concentration that resulted in a slower scintillator decay time. This is the first time that event-by-event direction reconstruction in high light-yield liquid scintillator has been demonstrated in a large-scale detector. 
\end{abstract}

\maketitle

\section{\textbf{\textit{Introduction}}} Organic liquid scintillator (LS) detectors play a central role in particle physics, particularly in studies of neutrinos, where numerous breakthrough measurements have been made in areas such as solar neutrinos \cite{Borexino_Nature}, reactor antineutrinos \cite{KamLAND:2002uet, DayaBay:2012fng,RENO:2012mkc,DoubleChooz:2012gmf} and neutrinoless double beta decay \cite{KamLAND-Zen:2022tow}. Several large scale LS detectors are currently in operation \cite{Borexino,Ozaki:2019uyd,SNO:2021} with more under construction \cite{JUNO} or being planned \cite{THEIA,Beacom_2017}. 

There has been much interest in recent years in enhancing the capabilities of such detectors by separating Cherenkov light from the scintillation signal so as to provide directional information while maintaining the energy resolution of high light-yield scintillators \cite{Biller:2013wua,Aberle:2014,Li:2015phc,Guo:2017nnr,YEH201151,Gruszko_2019,PhysRevD.101.072002,Kaptanoglu_2019,Caravaca_2017,PhysRevD.103.052004,Bonventre:2018hyd,Kaptanoglu_2022,Dunger2022}. Several recent community-planning exercises in both the U.S. and Europe have highlighted the importance of these developments \cite{huber2022snowmass,klein2022future,EFCA2021,acharya2023fundamental}. This concept is central to several small-scale demonstrators both under construction and in operation \cite{Anderson_2023,Annie2021,Svoboda2023}, as well as large-scale experiments currently under development \cite{THEIA,Beacom_2017}. Detectors have used Cherenkov light before in low-light yield scintillators, such as Liquid Scintillator Neutrino Detector (LSND), which had an isotropic to Cherenkov light ratio of 5:1 \cite{LSND1997} and MiniBooNE, which had a ratio of 3:1 \cite{Shaevitz_2008,MiniBooNE2009}. However, the Cherenkov component in high light-yield LS detectors typically only represents a few percent of the overall light signal, posing a considerably greater challenge. In addition to having a different wavelength profile, the prompt and directional nature of this light offers potential handles to distinguish it from the slower, isotropic scintillation signal. This can be aided by further slowing the characteristic scintillation time either by introducing primary fluors with longer intrinsic time constants \cite{Biller:2013wua,SlowFluors} or by reducing the primary fluor concentration to reduce the nonradiative coupling with the solvent \cite{Li:2015phc, Guo:2017nnr}.

During its commissioning phase, the SNO+ experiment operated with an initial concentration of \SI{0.6}{\gl} 2,5-diphenyloxazole (PPO) in the linear alkylbenzene (LAB) solvent, resulting in a scintillation decay timescale of $>$10 ns. Data during this time period shows a clear directional signal from $^8$B solar neutrino interactions. While a recent study by the Borexino Collaboration \cite{Borexino_direction,basilico2023final} found a correlation with the solar direction for individual early Cherenkov photons from a few percent of their solar events, and used those integrated distributions in fits to help extract signals, this is the first time that a large-scale experiment has demonstrated event-by-event direction reconstruction in high light-yield liquid scintillator.

\section{\textbf{\textit{The SNO+ Detector and Data Selection}}} SNO+ is a multipurpose neutrino detector located \SI{2}{\km} underground at SNOLAB in Ontario, Canada. Much of the infrastructure has been repurposed from the SNO experiment, including 9362 photomultiplier tubes (PMTs) and a \SI{6}{\m} radius acrylic vessel (AV). Significant upgrades have been made to the detector, allowing for the AV to be filled with LS \cite{SNO:2021LS}. Further details of the SNO+ detector can be found in \cite{SNO:2021}.

In order to select a pure sample of \ce{^8B} solar neutrinos for the current study, an energy region for the scattered electron of approximately \SIrange{5}{15}{\mev} was used \cite{SNO:2015wyx}. This removes the vast majority of events due to Uranium and Thorium chain (U/Th) chain backgrounds as well as events associated with muons. The higher energy events in this sample are expected to provide better directional information, compared to those below \SI{5}{\mev}, due to the increased number of Cherenkov photons, reduced electron multiple scattering and neutrino interaction kinematics. A dead time of \SI{20}{\s} was also enforced after any energy deposition above \SI{15}{\mev} in order to remove any further activity following muon events. A fiducial radius of \SI{5.5}{\m} from the center of the AV was used to select events from a region of more uniform detector optical response. After cuts, the only expected background events result from atmospheric neutrinos and the decays of $^{208}$Tl and $^{210}$Tl, which results in a total of $\sim1$ expected events within the dataset.

\section{\textbf{\textit{Datasets}}} 
\begin{figure}[H]
\begin{center}
\includegraphics[width = 0.95\linewidth]{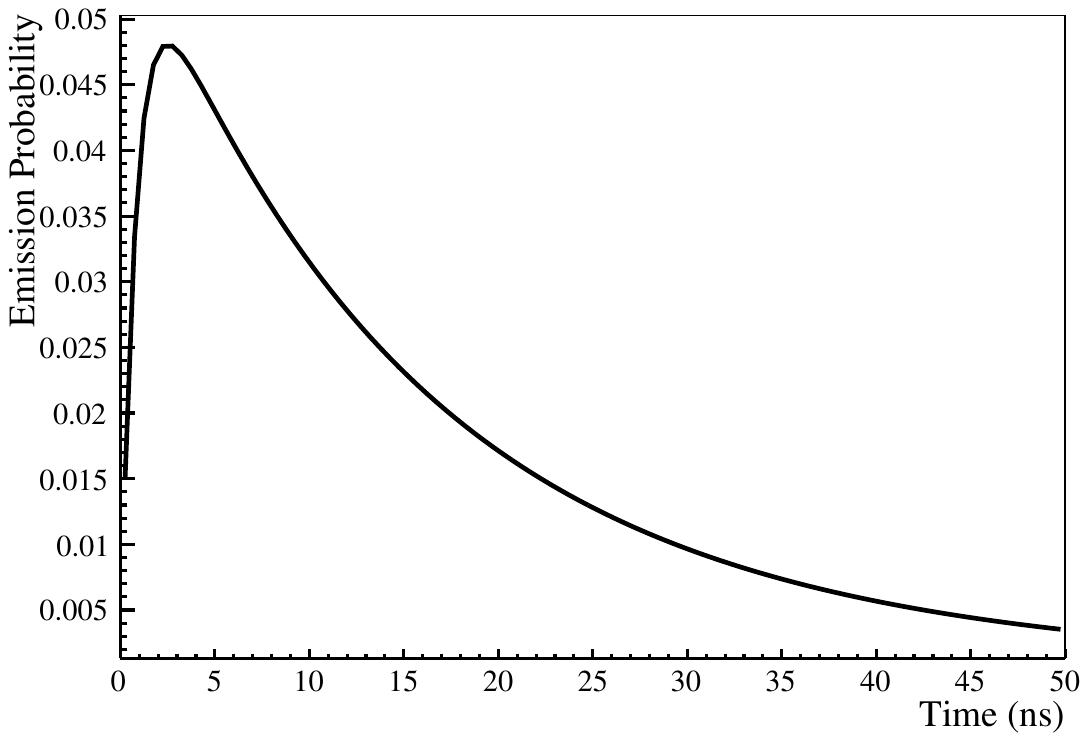}
\caption{Parameterized time profile of scintillation light emission for electrons in LAB with \SI{0.6}{\gl} PPO based on comparisons of $^{214}$Bi background events in data and simulations.\vspace{-0.3cm}}
\label{fig:ScintProfiles}
\end{center}
\end{figure}

During the period of April -- October 2020 (Period 1, exposure \SI{23}{\kilo\tonne\day}), SNO+ was partially filled with scintillator that floated on a larger volume of water inside the AV. The LS portion comprised  \SI{365}{\tonne} with a PPO concentration of \SI{0.6}{\gl}. The interface with the water region was  $\sim$\SI{75}{\cm} above the AV equator. {\it In situ} measurements of light emitted from low energy background events in the LS region during this phase indicated a yield of $\sim$300 detected photoelectrons per MeV of deposited electron energy. A parametrization of the inherent scintillator timing profile was derived based on a comparison of data with simulations for $^{214}$Bi background events, tagged via the associated alpha decay of $^{214}$Po, and has the following form:
\begin{equation}
    P(t) = \sum_{i=1}^3 A_i \frac{e^{-t/\tau_i} - e^{-t/\tau_r}}{\tau_i - \tau_r}
\end{equation}
where $P$ is the probability of photon emission, t is the time of emission, $A_i$ is the fraction of light in the ith component, $\tau_i$ is the fall time of the ith component and $\tau_r$ is the common rise time, found to be \SI{0.8}{\nano\s}. This parameterization is shown in Fig. \ref{fig:ScintProfiles} and the derived constants are given in Table \ref{tab:TimeConstants}.

\begin{table}[H]
\begin{center}
\begin{tabular}{|c|c|c|c|}
\hline
i & 1 & 2 & 3 \\
\hline
$\tau_i$ (ns) &13.5 & 23&98.5\\
$A_i$ &0.55 & 0.335&0.115\\
\hline
\end{tabular}
\end{center}
\caption{The derived scintillation timing profile parameters of LAB with \SI{0.6}{\gl} PPO.\vspace{-0.3cm}}
\label{tab:TimeConstants}
\end{table}%

For the Period 1 dataset, an extra fiducial volume constraint was implemented in order to avoid proximity to the scintillator-water interface, where optical effects became nonuniform. Events were excluded which reconstructed within the water volume, and within \SI{25}{\centi\m} above the interface. From this dataset, 20 solar neutrino candidates were extracted.

Between April and June 2021 (Period 2, exposure \SI{15}{\kilo\tonne\day}) the detector was filled with \SI{780}{\tonne} of LS at \SI{0.6}{\gl} PPO concentration. The scintillator optical, timing and light-yield characteristics were confirmed to be unchanged from Period 1. Seventeen solar neutrino candidates were extracted from this period, which were then combined with Period 1 to create a single dataset. 

The total dataset had an exposure of \SI{38}{\kilo\tonne\day} and 37 selected events, which is statistically consistent with the expectation of 40 events, derived from the flux presented in \cite{RevModPhys.92.045006}.

\section{\textbf{\textit{Cherenkov Separation in SNO+}}}
\begin{figure}[H]
\begin{center}
\begin{subfigure}{0.45\linewidth}
\includegraphics[width=\linewidth]{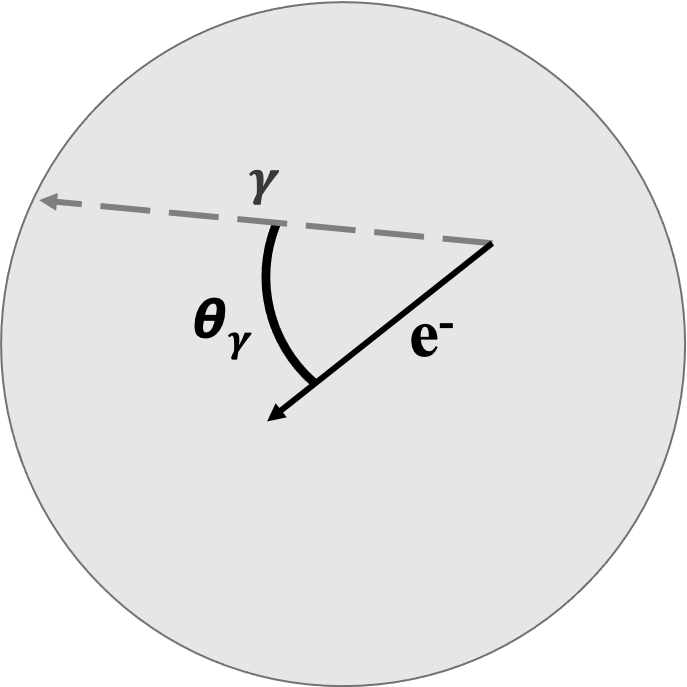}
\caption{}
\label{subfig:ThetaGamma}
\end{subfigure}
\begin{subfigure}{0.45\linewidth}
\includegraphics[width=\linewidth]{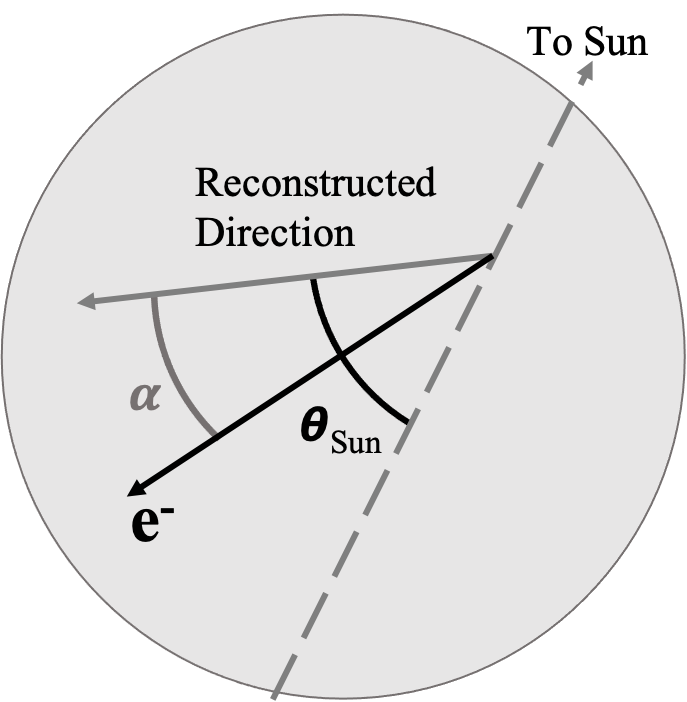}
\caption{}
\label{subfig:ThetaSun_Alpha}
\end{subfigure}
\caption{Definitions of angles referenced in this paper: $\theta_{\gamma}$ is the angle between the electron's travel direction and the photon direction; $\alpha$ is the angle between the true and reconstructed direction of the electron; and $\theta_{\text{Sun}}$ is the angle between the reconstructed event direction and the solar direction (assumed true direction of the neutrino).\vspace{-0.3cm}}
\label{fig:Angles}
\end{center}
\end{figure}

\vspace{-0.3cm} Reconstruction of event direction relies on isolating the instantaneous Cherenkov light from the scintillation signal using timing information. A ``time residual" is defined as:
\begin{equation}
t_{\text{res}} = t_{\text{hit}} - t_{\text{event}} - t_{\text{flight}}
\label{eqtres}
\end{equation}
where $t_{\text{hit}}$ is the recorded hit time of the PMT, $t_{\text{event}}$ is the reconstructed event time and $t_{\text{flight}}$ is the estimated time of flight of the photon, assuming a straight line light path. The latter two terms result from a maximum likelihood fit to an assumed pointlike vertex position based on timing information from all hit PMTs in the event. 
The distribution of $t_{\text{res}}$ is largely dominated by the inherent time spectrum of scintillation/Cherenkov light, the PMT time response, and uncertainties in the reconstructed vertex position. Within the selected energy range, the 3D vertex resolution is \SIrange{24}{26}{\centi\m}. The anisotropy in the early light due to the Cherenkov component can be identified by using this timing information in conjunction with the parameter $\theta_{\gamma}$, defined as the angle between the estimated photon direction ({\em i.e.} from the reconstructed vertex to the hit PMT) and the initial direction of the original electron, as shown in Fig. \ref{subfig:ThetaGamma}.

\begin{figure}[H]
\begin{center}
\includegraphics[width=0.95\linewidth]{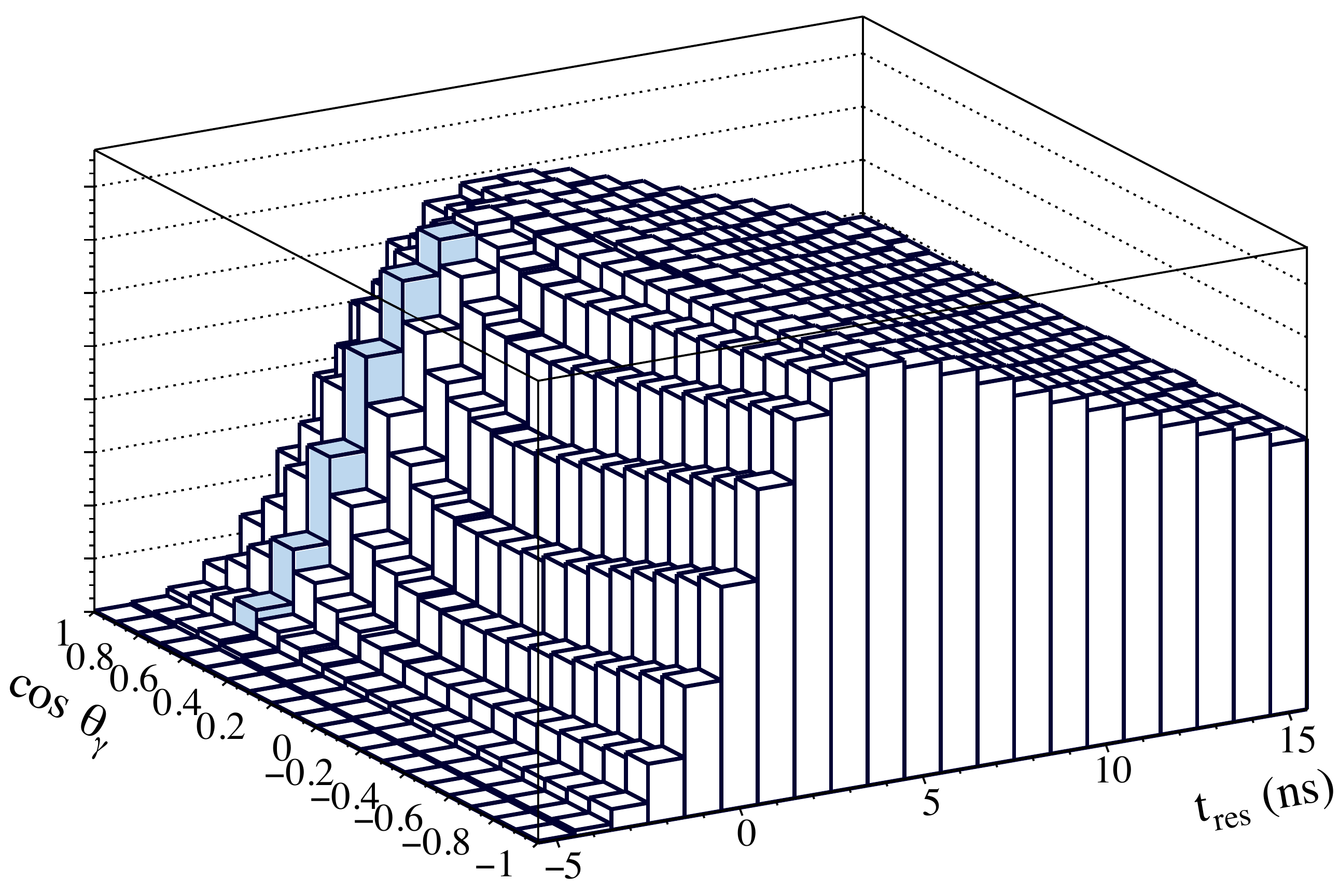}
\caption{Distribution of photon hits in $\cos\theta_{\gamma}$ and $t_{\text{res}}$ for simulated \SI{6}{\mev} electrons in LAB with a PPO concentration of \SI{0.6}{\gl}. A clear peak can be seen at low $t_{\text{res}}$ near the expected Cherenkov angle, $\cos\theta_{\gamma} = 0.66$, highlighted in blue.\vspace{-0.3cm}}
\label{fig:PDF}
\end{center}
\end{figure}

It is possible to clearly see this  anisotropy in the $\cos\theta_{\gamma}$ distribution on the rising edge of the scintillator timing profile in simulations of \SI{6}{\mev} electrons using measured scintillation characteristics, where \SI{6}{\mev} was chosen to represent a typical electron recoil energy from a $^8$B solar neutrino for this dataset. A clear Cherenkov peak can be seen in Fig. \ref{fig:PDF} at low $t_{\text{res}}$ near the expected emission angle of $\cos\theta_\gamma=0.66$. There is also a slight bias in the  ``backwards" direction (towards $\cos \theta_\gamma = -1$) evident in the plot. This is due to bias in the vertex reconstruction caused by the earlier Cherenkov photons ``pulling" the best fit vertex along the direction of motion. This effect was also noted in Borexino's directionality studies \cite{Borexino_direction,basilico2023final}. This suggests that it might be possible to improve reconstruction by simultaneously fitting for both vertex and direction. While this is currently under study, the present analysis treats these two aspects separately.

\section{\textbf{\textit{Direction Reconstruction}}} 
\begin{figure}[H]
\begin{center}
\includegraphics[width=0.95\linewidth]{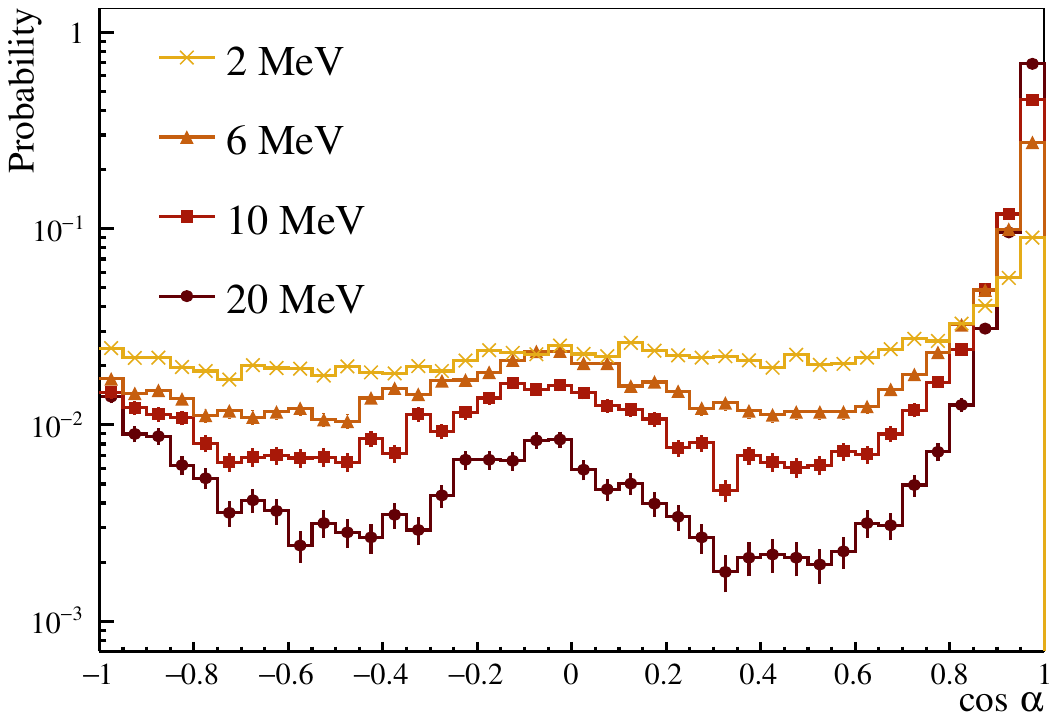}
\caption{Results of direction reconstruction of simulated electrons of different energies, with error bars displaying statistical uncertainties.\vspace{-0.3cm}}
\label{fig:Energies}
\end{center}
\end{figure}

The distribution in Fig. \ref{fig:PDF} is used as a probability density function (PDF) for a maximum likelihood reconstruction of event direction. Results of this direction reconstruction for electrons of different energies simulated in the SNO+ detector are shown in Fig. \ref{fig:Energies}, where $\alpha$ is defined as the angle between the true direction and the reconstructed direction as shown in Fig. \ref{subfig:ThetaSun_Alpha}. Table~\ref{tab:Cos_Energies} shows simulation predictions for the percentage of events with $\cos\alpha > 0.8$ for different electron energies.

As shown in Fig. \ref{fig:Energies}, higher energy electrons yield better direction reconstruction owing to increased photon sampling as well as a reduced impact from electron multiple scattering. Increased effective photocathode coverage and/or slower scintillator formulations should be able to extend the usable range of direction reconstruction to lower energies \cite{Dunger2022}. 

\begin{figure}[H]
\begin{center}
\includegraphics[width=0.95\linewidth]{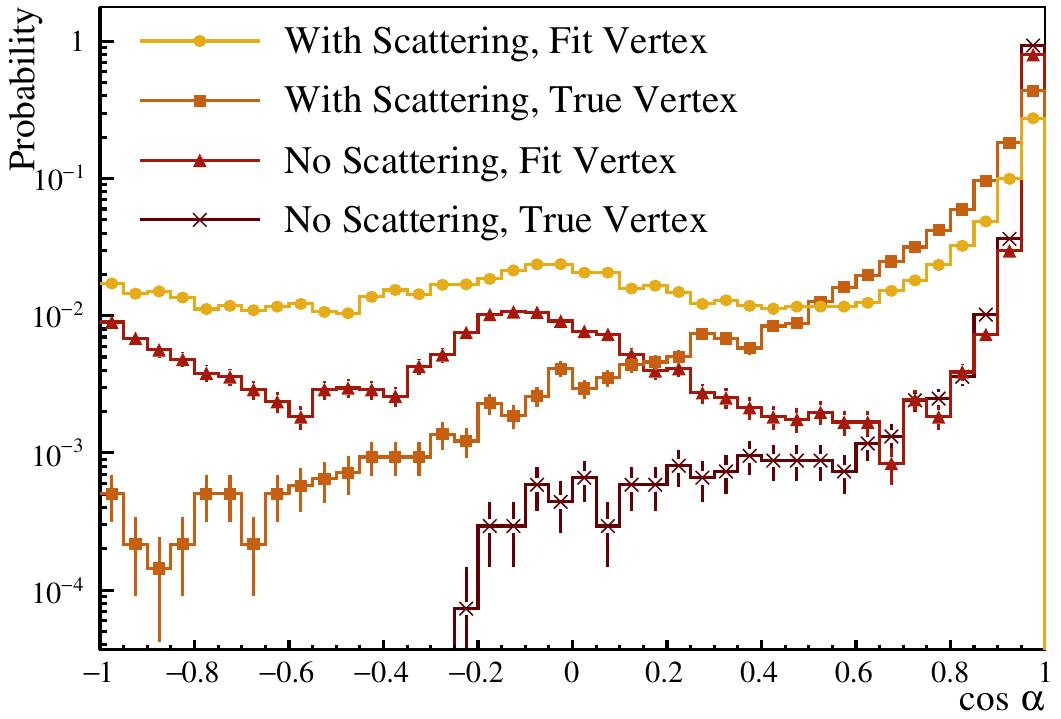}
\caption{Direction reconstruction of simulated \SI{6}{MeV} electrons in scintillator for cases of 1) no multiple scattering and using the true vertex position (brown, cross); 2) no multiple scattering and using the fit vertex position (red, triangle); 3) multiple scatterings and using the true vertex position (orange, square); and 4) multiple scatterings and using the fit vertex position (yellow, circle).\vspace{-0.3cm}}
\label{fig:Effects}
\end{center}
\end{figure}

The impact of multiple scattering and vertex reconstruction bias on the direction fit are illustrated in Fig. \ref{fig:Effects} for simulated \SI{6}{\mev} electrons. From this investigation, it is clear the features near $\cos\alpha=0$ and $\cos\alpha=-1$ are caused by vertex reconstruction effects.

\begin{table}[H]
\begin{center}
\begin{tabular}{|c|c|}
\hline
Electron Energy (MeV)& {\% with $\cos\alpha > 0.8$} \\
\hline
2 & $21.9 \pm 0.4$ \\
6 & $45.6 \pm 0.6$ \\
10 & $64.6 \pm 0.7$ \\
20 & $83.0 \pm 0.8$ \\
\hline
\end{tabular}
\end{center}
\caption{Energy dependence of direction reconstruction for electrons simulated in the SNO+ detector, with statistical uncertainties.  The performance is quantified by the percentage of events that reconstruct with $\cos\alpha > 0.8$.\vspace{-0.3cm}}
\label{tab:Cos_Energies}
\end{table}%

\section{\textbf{\textit{Application to Solar Neutrinos}}} 
\begin{figure}[H]
\vskip 0.2in
\begin{center}
\includegraphics[width=0.95\linewidth]{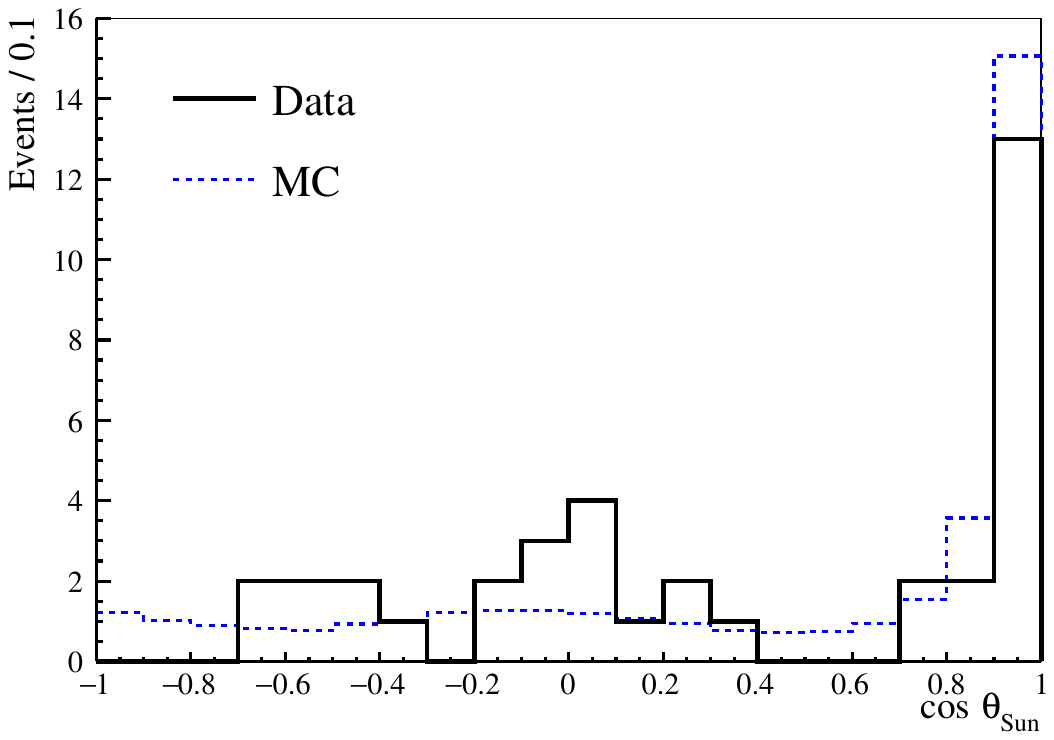}
\caption{Results of direction reconstruction for measured (solid) and simulated (dashed) $^8$B solar neutrinos, where simulation sampled from a nominal $^8$B energy spectrum.\vspace{-0.3cm}}
\label{data}
\end{center}
\end{figure}

Direction reconstruction was applied to the solar neutrino dataset previously described, with 37 selected events containing $\sim 1$ expected background event. Due to the kinematics of solar neutrino elastic-scattering, the scattered electrons have additional angular spread relative to the solar direction, as indicated in Fig. \ref{subfig:ThetaSun_Alpha}. Due to the selection of electrons with an energy greater than \SI{5}{\mev}, this smearing is confined to one bin. The results of direction reconstruction in data are shown in Fig. \ref{data}, along with Monte Carlo (MC) predictions sampled from a nominal $^8$B energy spectrum. A clear peak at $\cos\theta_{\text{Sun}} = 1$ can be seen. A likelihood ratio test in comparison with an isotropic distribution yields a $p$-value corresponding to 5.7$\sigma$.

\section{\textbf{\textit{Conclusions}}} Data from the SNO+ experiment has been used to demonstrate event-by-event direction reconstruction of recoil electrons from solar neutrinos for the first time, exploiting the time separation of the directional Cherenkov light from the scintillation light. Remarkably, this demonstrates the ability to achieve comparable directional information to that of water Cherenkov detectors, which has played a crucial role in numerous fundamental measurements. This new capability opens up interesting possibilities for current detectors and for the design of future instruments. Potential applications include studies of solar neutrinos, supernova neutrinos, and background discrimination for other physics. By utilizing slower scintillators, increased photocathode coverage and/or other developing technologies, this approach could be extended to lower energies, where it could also provide important background suppression (such as from $^8$B solar neutrinos) in studies of phenomena such as neutrinoless double beta decay.

Capital funds for SNO+ were provided by the Canada Foundation for Innovation and matching partners: Ontario Ministry of Research, Innovation and Science, Alberta Science and Research Investments Program, Queen’s University at Kingston, and the Federal Economic Development Agency for Northern Ontario. This research was supported by (Canada) the Natural Sciences and Engineering Research Council of Canada, the Canadian Institute for Advanced Research, the Ontario Early Researcher Awards; (U.S.) the Department of Energy (DOE) Office of Nuclear Physics, the National Science Foundation, and the DOE National Nuclear Security Administration through the Nuclear Science and Security Consortium; (U.K.) the Science and Technology Facilities Council and the Royal Society; (Portugal) Fundação para a Ciência e a Tecnologia (FCT-Portugal); (Germany) the Deutsche Forschungsgemeinschaft; (Mexico) DGAPA-UNAM and Consejo Nacional de Ciencia y Tecnología; and (China) the Discipline Construction Fund of Shandong University. We also thank SNOLAB and SNO+ technical staff; the Digital Research Alliance of Canada; the GridPP Collaboration and support from Rutherford Appleton Laboratory, and the Savio computational cluster at the University of California, Berkeley. Additional long-term storage was provided by the Fermilab Scientific Computing Division. 

For the purposes of open access, the authors have applied a Creative Commons Attribution licence to any Author Accepted Manuscript version arising. Representations of the data relevant to the conclusions drawn here are provided within this paper.

\bibliography{DirectionPaper}

\end{document}